\documentclass[prl,aps,showpacs,twocolumn,floatfix]{revtex4-1}
\usepackage{amsmath,amssymb,graphicx,cases}
\usepackage{epsfig}
\usepackage{textcomp}
\usepackage{epstopdf}
\usepackage{float}
\usepackage[normalem]{ulem}
\newcommand{\stkout}[1]{\ifmmode\text{\sout{\ensuremath{#1}}}\else\sout{#1}\fi}

\usepackage[colorlinks=true, urlcolor=blue, anchorcolor=blue, citecolor=blue,filecolor=blue,linkcolor=blue,menucolor=blue,pagecolor=blue]{hyperref}

\begin{document}
\title{Comment on ``Minimum Action Path Theory Reveals the Details of Stochastic Transitions
Out of Oscillatory States" }

\author{Baruch Meerson}
\email{meerson@mail.huji.ac.il}
\affiliation{Racah Institute of Physics, Hebrew University of
Jerusalem, Jerusalem 91904, Israel}
\author{Naftali R. Smith}
\email{naftalismith@gmail.com}
\affiliation{Racah Institute of Physics, Hebrew University of
Jerusalem, Jerusalem 91904, Israel}


\maketitle

In a recent Letter \cite{Carrasco} de la Cruz \textit{et al}.  studied a noise-induced transition in an oscillating stochastic population undergoing birth- and death-type reactions. When described by deterministic rate equations, the population approaches a stable limit cycle.
The intrinsic noise, caused by the discreteness of molecules and randomness of their interactions, leads to escape from this limit cycle through an adjacent unstable limit cycle, and  de la Cruz \textit{et al}. attempted to evaluate the mean first passage time (MFPT) to escape.

A crucial approximation, made in the Letter, was to replace the original Master equation by the ``chemical Langevin equation" (CLE), their Eq. (2). Unfortunately, this standard procedure, based on the van Kampen expansion  in the inverse population size $1/\Omega \ll 1$ \cite{vK},  applies only for typical, small fluctuations around the stable limit cycle. It fails in the tails of the metastable quasi-stationary distribution of the population size around the limit cycle. One of these tails determines
the escape rate of the population through the unstable limit cycle. As a result, the MFPT, predicted by de la Cruz \textit{et al}, involves an error which grows exponentially with the population size $\Omega \gg 1$, due to an error in the calculation of $\mathcal{S}$. In this situation their study of a pre-exponential factor in the MFPT does not have much meaning.

The inadequacy of the van Kampen system-size expansions for a description of large fluctuations in Markov jump processes is by now well documented \cite{Gaveau,EK,Doering,KS,AM2007,OM,AM2017}. The only general exception appears when the system is close to the proper bifurcation of the underlying deterministic model \cite{Doering,MS2008,EK2009,AM2010,AM2017}. In the present case it is the saddle-node bifurcation of the stable and unstable limit cycles.

Fortunately, there is no need for uncontrolled approximations. The Freidlin-Wentzell WKB theory was extended to stochastic populations quite some time ago \cite{Kubo,Gang,Dykman}. The corresponding WKB technique employs the same large parameter $\Omega\gg 1$ but circumvents the van Kampen system-size expansion, see\textit{ e.g.} Ref. \cite{AM2017} for a recent review. Moreover, this WKB technique was already applied to escape from a limit cycle, in the context of extinction of long-lived oscillating populations \cite{SM}.

Even within the framework of the CLE, much of the Letter is devoted to a rediscovery of known results, as de la Cruz \textit{et al}. seem to be unaware of a body of important previous analytical, numerical and experimental work on noise-induced escape from limit cycles and from  attractors of dynamical systems in general \cite{Kautz1987,Kautz1988,Grassberger,Kautz1996,Smelyanskiy1997,Khovanov2000}. A proper formulation of the Freidlin-Wentzell escape optimization problem, which was put forward in these works, and which is lacking in the Letter,  involves the time interval $-\infty<t<\infty$.  A minimum action path -- an instanton -- exits the limit cycle at $t=-\infty$ while performing an infinite number of loops. There is a whole one-parameter family of instanton solutions, linked to one another through the time translations $t\to t+ \text{const}$, and each instanton yields the same classical action. Any evidence to the contrary results from finite-time numerical artifacts.
\vspace{0.1cm}

We acknowledge discussions with M.I. Dykman and support from the Israel Science Foundation (Grant No. 807/16).


\begin{thebibliography} {99}
\bibitem{Carrasco} R. de la Cruz, R. Perez-Carrasco, P. Guerrero, T. Alarcon, and K. M. Page, Phys. Rev. Lett. \textbf{120}, 128102 (2018);  arXiv:1705.08683.
\bibitem{vK} N. van Kampen, \textit{Stochastic Processes in Physics and Chemistry}
(Elsevier, New York, 2007).
\bibitem{Gaveau} B. Gaveau, M. Moreau, and J. Toth, Lett. Math. Phys. \textbf{37}, 285 (1996).
\bibitem{EK} V. Elgart and A. Kamenev, Phys. Rev. E 70, 041106
(2004).
\bibitem{Doering} C. R. Doering, K.V. Sargsyan, and L. M. Sander,
Multiscale Model. Simul. 3, 283 (2005).
\bibitem{KS} D. A. Kessler and N. Shnerb, J.  Stat. Phys. \textbf{127}, 861 (2007).
\bibitem{AM2007} M. Assaf and B. Meerson, Phys. Rev E \textbf{75}, 031122 (2007).
\bibitem{OM} O. Ovaskainen and B. Meerson, Trends Ecol. Evol. \textbf{25}, 643 (2010).
\bibitem{AM2017} M. Assaf and B. Meerson, J. Phys. A: Math. Theor. \textbf{50}, 263001 (2017).
\bibitem{MS2008} B. Meerson and P.V. Sasorov, Phys.  Rev.  E, \textbf{78}, 060103(R) (2008).
\bibitem{EK2009} C. Escudero and A. Kamenev, Phys. Rev. E \textbf{79}, 041149 (2009).
\bibitem{AM2010} M. Assaf and B. Meerson, Phys. Rev E \textbf{81}, 021116 (2010).
\bibitem{Kubo} R. Kubo, K. Matsuo, and K. Kitahara K J. Stat. Phys. \textbf{9}, 51 (1973).
\bibitem{Gang} H. Gang, Phys. Rev. A \textbf{36}, 5782 (1987).
\bibitem{Dykman} M. I. Dykman, E. Mori, J. Ross, and P. M. Hunt, J. Chem. Phys
\textbf{100}, 5735 (1994).
\bibitem{SM} N.R. Smith and B. Meerson, Phys. Rev. E \textbf{93}, 032109 (2016).
\bibitem{Kautz1987} R. Kautz, Phys. Lett. A \textbf{125}, 315 (1987).
\bibitem{Kautz1988} R. Kautz, Phys. Rev. A \textbf{38}, 2066 (1988).
\bibitem{Grassberger} P. Grassberger, J. Phys. A: Math. Gen. \textbf{22}, 3283 (1989).
\bibitem{Kautz1996} R.L. Kautz,  Rep. Prog. Phys. \textbf{59}, 935 (1996).
\bibitem{Smelyanskiy1997} V. N. Smelyanskiy, M. I. Dykman, and R. S. Maier, Phys. Rev. E
\textbf{55}, 2369 (1997); 
Phys. Rev. E \textbf{56}, 2332 (1997).
\bibitem{Khovanov2000} I. A. Khovanov, D. G. Luchinsky, R. Mannella, and P.
V.E. McClintock, in ``Stochastic Processes in Physics,
Chemistry and Biology", edited by J. A. Freund and T. P\"{o}schel (Springer, Berlin, 2000), p. 378.



\end{thebibliography}
\end{document}